\documentclass[12pt,letter,preprint]{elsarticle}

\usepackage{lineno,hyperref}
\modulolinenumbers[5]
\usepackage{cancel}

\journal{Physics Letters B}

\usepackage{xcolor}

\usepackage{fancyhdr}
\fancypagestyle{titlepage}{

\lhead{}
\rhead{PITT-PACC-1820}
}

\begin{document}

\begin{frontmatter}

\title{SUSY Signals from QCD Production at the Upgraded LHC}

\author{Tao Han}
\ead{than@pitt.edu}
\author{Ahmed Ismail}
\ead{aismail@pitt.edu}
\author{Barmak Shams Es Haghi}
\ead{bas143@pitt.edu}
\address{PITT-PACC, Department of Physics and Astronomy, University of Pittsburgh, Pittsburgh, PA 15260, USA}

\begin{abstract}
Weak-scale supersymmetry remains to be one of the best-motivated theories of physics beyond the Standard Model. We evaluate the sensitivities of the High Luminosity (HL) and High Energy (HE) upgrades of the LHC to gluinos and stops, decaying through the simplified topologies $\tilde{g} \to q \bar{q} \chi^0$, $\tilde{g} \to t \bar{t} \chi^0$ and 
$\tilde{t} \to t \tilde{\chi}^0$. Our HL-LHC analyses improve on existing experimental projections by optimizing the acceptance of kinematic variables. The HE-LHC studies represent the first 27 TeV analyses. We find that the HL-(HE-)LHC with 3 ab$^{-1}$ (15 ab$^{-1}$) of integrated luminosity will be sensitive to the masses of gluinos and stops at 3.2 (5.7) TeV and 1.5 (2.7) TeV, respectively, decaying to massless neutralinos.
\end{abstract}

\end{frontmatter}

\thispagestyle{titlepage}

\section{Introduction and Methodology}
\label{sec:intro}

As one of the leading resolutions of the hierarchy problem associated with the weak scale and the Planck scale, supersymmetry (SUSY) has attracted enormous attention as an experimental target for past, present, and future colliders. The lamppost of naturalness suggests that super-partners should appear near the electroweak scale \cite{Feng:2013pwa}. Although experimental searches have not established any observable signal thus far, it is conceivable that the SUSY scale may still be just out of reach or that SUSY experimental signatures are unconventional in nature. With this in mind, it is essential to develop and quantify the impact of SUSY search strategies at the Large Hadron Collider (LHC) and its potential successors.

At the end of Run 2 of the LHC, preparations for the luminosity upgrade of the LHC are well underway \cite{Apollinari:2015bam}. In addition, there has been significant interest in the possibility of augmenting the LHC energy with new stronger magnets \cite{Zimmermann:2018wdi}. The energy upgrade of the LHC could take us to the next energy frontier. We thus find it timely to reexamine SUSY searches. In this paper, we describe the ability of luminosity- and energy-upgraded versions of the LHC to probe supersymmetry.

Supersymmetric theories are broad in scope, and even if one is to focus solely on collider searches, there is much model-dependence in signature space~(see, {\it e.g.},~\cite{Cahill-Rowley:2014twa}). 
A variety of search strategies are necessary to cover the multitude of potential signatures of SUSY. In this work we consider two well-motivated and leading channels, namely, pair production of the gluinos and 
top-quark partners (stops). Once kinematically accessible, gluino pair production will be the leading channel because of its octet representation under QCD.
On the other hand, owing to the large top-quark Yukawa coupling and the resulting stop soft-mass evolution over scales, it is motivated to consider the stops as the lightest quark partners and thus the most kinematically favored for production.
Specifically, we evaluate the leading decay modes of hadronic jets plus large missing transverse energy ($\cancel{E}_T$), resulting from the missing neutral lightest SUSY partner (LSP), taken to be the neutralino ($\chi^0$). We also include flavor tagging, requiring that some of the jets be $b$-tagged. 
Thus the standard decay chains of gluinos and stops under consideration are 

$$\tilde{g} \to q \bar{q} \chi^0,\ \tilde{g} \to t \bar{t} \chi^0, \ {\rm and}\ \tilde{t} \to t \chi^0.$$ 

From a bottom-up perspective, the stop and gluino are important to search for because they affect the Higgs mass parameter at one and two loops respectively, and with their strong couplings are expected to be some of the most important super-particles with respect to tuning of the electroweak scale\footnote{The Higgsino affects the Higgs mass at tree level, but is uncolored and thus more challenging to produce at a hadron collider. We note that while we consider models with bino LSPs, there is an implicit tuning at large LSP mass because the Higgsino must be even heavier than the LSP.}. Throughout, we consider simplified models~\cite{Alwall:2008va,Alwall:2008ag} wherein the new particle production is entirely due to gluinos and stops which decay through the above modes to a bino LSP~\cite{Baer:1990sc,Baer:1994xr}.
We note that while the analyses which we consider are some of the most common collider SUSY searches, their relevance depends on SUSY breaking scenarios and parameter assumptions, including the mass splitting between the colored super-partners and the LSP. Nevertheless, they give a representative starting point and a perspective for well-motivated channels.

We estimate the potential reach of these searches at both the High Luminosity LHC (HL-LHC), with 3 ab$^{-1}$ of 14 TeV proton collisions, and the High Energy LHC (HE-LHC), with the energy and integrated luminosity increased to 27 TeV and 15 ab$^{-1}$, respectively. 
While over the past several years HL-LHC and 100 TeV analyses have sought to project the potential of future colliders to search for the same superparticles as we consider (see, e.g.,~\cite{Gershtein:2013iqa,CEPC-SPPCStudyGroup:2015csa,Arkani-Hamed:2015vfh,Golling:2016gvc} for reviews), HE-LHC studies are less common (though some 33 TeV studies exist~\cite{Baer:2017pba}), and have considered different final states with varying motivations~\cite{Aboubrahim:2018bil,Baer:2018hpb}. This work is most closely related to the ATLAS HL-LHC studies~\cite{ATL-PHYS-PUB-2014-010,Aaboud:2016zdn}, which we have both improved upon with cut optimization and the inclusion of currently accepted detector projections, and extended to the HE-LHC. One of the gluino searches presented here has appeared in the report of the Beyond the Standard Model Physics Working Group~\cite{CidVidal:2018eel} in the \emph{Workshop on Physics at the HL-LHC, and perspectives on HE-LHC}, while the other extended results appear here for the first time. 

The rest of this paper is organized as follows. In Section~\ref{sec:gluinos}, we project the HL-LHC and HE-LHC gluino reach in the 4 jets plus $\cancel{E}_T$ final state, both in the cases where the gluino decays through light-flavor or heavy-flavor off-shell squarks. In Section~\ref{sec:stops}, we estimate the reach of the luminosity- and energy-upgraded LHC for the stop in the $b$-jets plus jets plus $\cancel{E}_T$ final state. Section~\ref{sec:concl} contains a discussion of our results and our conclusions.

\section{Gluinos}
\label{sec:gluinos}

While the gluino only affects the Higgs mass at two loops, it nevertheless plays an important numerical role in contributing to Higgs mass corrections \cite{Craig:2013cxa}, and enjoys a relatively large production cross section as a fermionic color octet. The total production cross section including NLO and NLL QCD corrections~\cite{Borschensky:2014cia} is, for gluinos of mass 2 TeV,
$${\rm 1.7\ fb \ at\ 14\ TeV\ and\ 68\ fb\ at\ 27\ TeV}.$$
The latter figure is estimated using the K-factor from 33 TeV gluino production, and shows an increase of a factor of 40 by going to the HE-LHC.
In our study, we evaluate the sensitivity of future proton colliders to gluino pair production with gluinos decaying to the LSP through off-shell squarks, using standard jets + $\cancel{E}_T$ searches. Given the simplified model we are considering as signal with SUSY masses as free parameters, we elect to optimize search regions requiring four jets and missing energy.

Before moving to the details of our studies, we comment briefly on our choice of final state. First, if light flavor squarks are kinematically accessible, $\tilde{q}\bar{\tilde{q}}$, $\tilde{q} \tilde{q}$, and $\tilde{q} \tilde{g}$ production would lead to events with only two or three hard jets. In addition, if a squark is present in the spectrum which weighs less than the gluino, the cascade decay $\tilde{g} \to q \tilde{q}, \tilde{q} \to q \tilde{\chi}^0$ can lead to somewhat different kinematics with a squark two-body decay, depending on the masses of the intermediate squark and the LSP. We have chosen the particular topology $\tilde{g} \to q \bar{q} \tilde{\chi}^0$ for two main reasons. First, from a bottom-up perspective there is less reason to expect the light flavor squarks to be very closely related to the weak scale, and indeed in
the limit of decoupled squarks final states with light-flavor quarks can be important. In addition, adding more particles such as intermediate squarks would increase the dimensionality of the parameter space to consider, and we prefer to avoid this complication here. Finally, in the compressed region where the gluino and LSP masses are similar, a search region with fewer jets is expected to be somewhat more effective~\cite{Cohen:2013xda}. We will nevertheless find that by loosening our cuts on $\cancel{E}_T$ and related kinematic variables, we can achieve reasonable sensitivity in this scenario. Typically, the four jets plus missing energy final state is the most powerful in constraining this simplified model~\cite{Aaboud:2017vwy}.

Similar considerations apply in the case of gluinos decaying through off-shell 3$^{\mathrm{rd}}$ generation squarks. It is useful to separate this topology both because of the motivation from naturalness for the stop to be close in mass to the electroweak scale, and because the $b$-quarks from top and bottom decays can be tagged, yielding different experimental signatures than for gluinos decaying through off-shell light flavor squarks. We thus choose to also consider the decay topology $\tilde{g} \to t \bar{t} \tilde{\chi}^0$. Due to the different decay modes of the tops, multiple final states can occur, including same-sign dileptons plus missing energy~\cite{Sirunyan:2017uyt,Aaboud:2017dmy}, but common to all of the possibilities are multiple $b$-jets. We will follow the authors of~\cite{Baer:2018hpb} and consider purely hadronic top decays, performing a search in a jets plus missing energy final state where leptons are vetoed and some of the jets are required to be tagged as originating from $b$ quarks. In principle there can be up to 12 hard jets in the final state when gluino decays proceed through off-shell stops and the resulting tops decay hadronically, but we consider only the first four jets in our analysis. We expect that with this choice, gluino decays through off-shell sbottoms will enjoy a similar limit.

\subsection{$\tilde{g} \to q \bar{q} \tilde{\chi}^0$}

The main SM backgrounds in the $4j + \cancel{E}_T$ final state are $Z (\to \nu \nu)$ + jets, $W (\to \ell \nu)$ + jets, and $t\bar{t}$ production. There are additional small contributions to the background from single top and diboson production, which we ignore. We generate signal and background using MadGraph 5~\cite{Alwall:2014hca} at parton level and Pythia 8~\cite{Sjostrand:2014zea} to perform showering with MLM matching~\cite{Mangano:2006rw}. At parton level, the background is generated in bins of $H_T$, where $H_T$ is the scalar sum of the $p_T$ of all jets, to ensure sufficient statistics. We simulate the detector response using Delphes 3~\cite{deFavereau:2013fsa}, which employs FastJet~\cite{Cacciari:2011ma} to cluster jets using the anti-$k_T$ algorithm~\cite{Cacciari:2008gp}, with the commonly accepted HL-LHC card corresponding to the upgraded ATLAS and CMS detectors. For the signal, we normalize the cross sections to NLL+NLO calculations~\cite{Borschensky:2014cia}. To encapsulate higher-order effects for background, we apply a universal K-factor of 1.25.

Following previous works~\cite{Cohen:2013xda,ATL-PHYS-PUB-2014-010,Aaboud:2016zdn}, we apply a set of baseline cuts
\begin{equation}
p_T (j_1)> 160~\mathrm{GeV},\,\, N_\mathrm{jets}(p_T > 60~\mathrm{GeV}\, ,|\eta| <5.0)\geq 4,\,\, \cancel{E}_T > 160~\mathrm{GeV} .
\label{eq:cut}
\end {equation}
We further require that signal events contain no isolated electrons (muons) with $p_T$ above 10 (10) GeV and $|\eta|$ below 2.47 (2.4). 
We reject events with $\Delta \phi(j, \cancel{E}_T)_{\mathrm{min}} < 0.4$, where $\Delta \phi(j, \cancel{E}_T)_{\mathrm{min}}$ is the minimum transverse angle between $\cancel{E}_T$ and the first three leading jets, to avoid contamination from QCD background with mismeasured jets. To further reduce the background, we demand $\cancel{E}_T / \sqrt{H_T}>10~\mathrm{GeV}^{1/2}$ and $p_T(\mathrm{j_4})/H_T > 0.1$, and ${\mathrm{j_4}}$ indicates the fourth leading jet. After this baseline selection, a two dimensional optimization over cuts on $\cancel{E}_T$ and $H_T$ is performed to obtain the maximum significance.
The latter cuts are optimized at each signal point to maximize the significance defined as
\begin{equation}
\mathcal{S} = {S \over \sqrt{B + \sigma_B^2 B^2+ \sigma_S^2 S^2}},
\end{equation}
where $S$ and $B$ are the number of signal and background events, respectively, and $\sigma_B = 20\%$  and $\sigma_S = 10\%$ are our assumed background and signal systematic uncertainties.
To assure a reasonable description of the statistical significance, we require at least 8 background events throughout our studies. 
For the HL-LHC (HE-LHC), we vary $\cancel{E}_T$ in steps of 0.5 (0.5) TeV from 0.5 (0.5) up to 3.0 (7.0) TeV and $H_T$ in steps of 0.5 (0.5) TeV from 0.5 (0.5) up to 5.0 (7.0) TeV.
 Because of our cut optimization, our search is predicted to enjoy better sensitivity than the existing ATLAS HL-LHC study~\cite{ATL-PHYS-PUB-2014-010}.

\begin{table*}
\centering
\begin{tabular}{|c|c|c|c|}
\hline
$m(\tilde{g}, \tilde{\chi}^0)$ & (1020, 1000)& (2000, 300)& Background\\
\hline \hline
Generator-level cuts &396 & 1.70  &$3.43 \cdot 10^4$\\
Lepton veto & 396 &1.69   &$1.93 \cdot 10^4$\\
Jet $p_T > (160, 60, 60, 60)$~GeV &30.8 & 1.64 & $5.45 \cdot 10^3$ \\
$\cancel{E}_T > 160$~GeV &  27.8&   1.59& $1.73 \cdot 10^3$ \\
$\Delta \phi(j_{1,2,3},\cancel{E}_T) > 0.4$ & 23.1 &1.27   & $1.10 \cdot 10^3$ \\
$\cancel{E}_T / \sqrt{H_T} > 10~\mathrm{GeV}^{1/2}$ & 18.6  & 0.97  &358 \\
$p_T(\mathrm{j_4})/H_T > 0.1$ &8.82  & 0.48 & 246\\
\hline \hline
$\cancel{E}_T > 500~\mathrm{GeV}$ & 2.82 & -- &17.6 \\
$H_T > 1000~\mathrm{GeV}$ & 1.65 & -- & 5.21\\
\hline
$\cancel{E}_T > 500~\mathrm{GeV}$ & -- &0.46   & 17.6\\
$H_T> 3500~\mathrm{GeV}$ & -- & 0.09  &0.004 \\
\hline
\end{tabular}
\caption{Cut flow for the two gluino benchmark points in Eq.~(\ref{eq:BP1}) at HL-LHC.
At each step, the cross section remaining after the indicated cut is shown in fb. The baseline cuts are identical for both points as stated in the text.}
\label{tab:gluinocutflow}
\end{table*}

\begin{figure}
\centering
\includegraphics[width=0.49 \textwidth ]{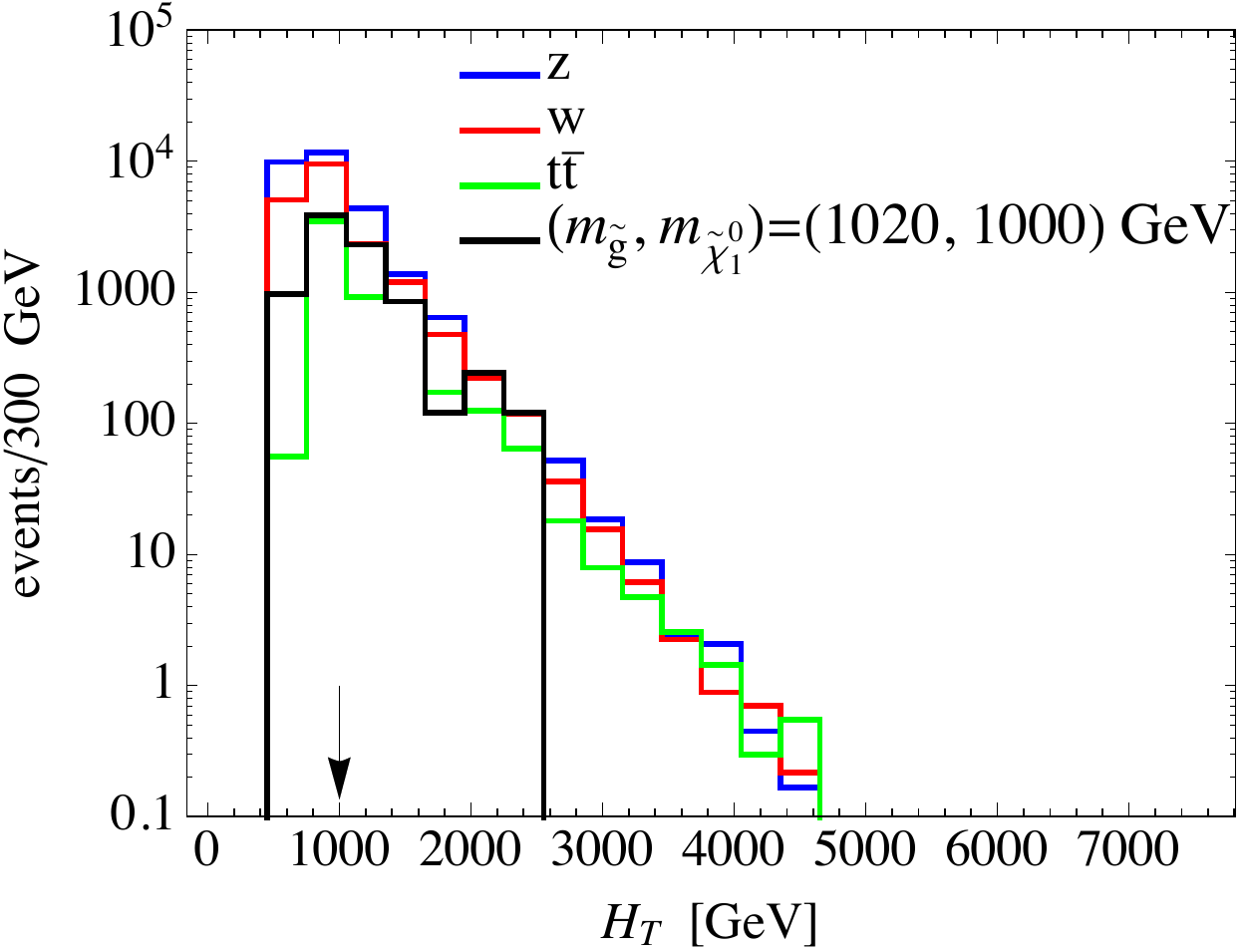}
\includegraphics[width=0.49 \textwidth ]{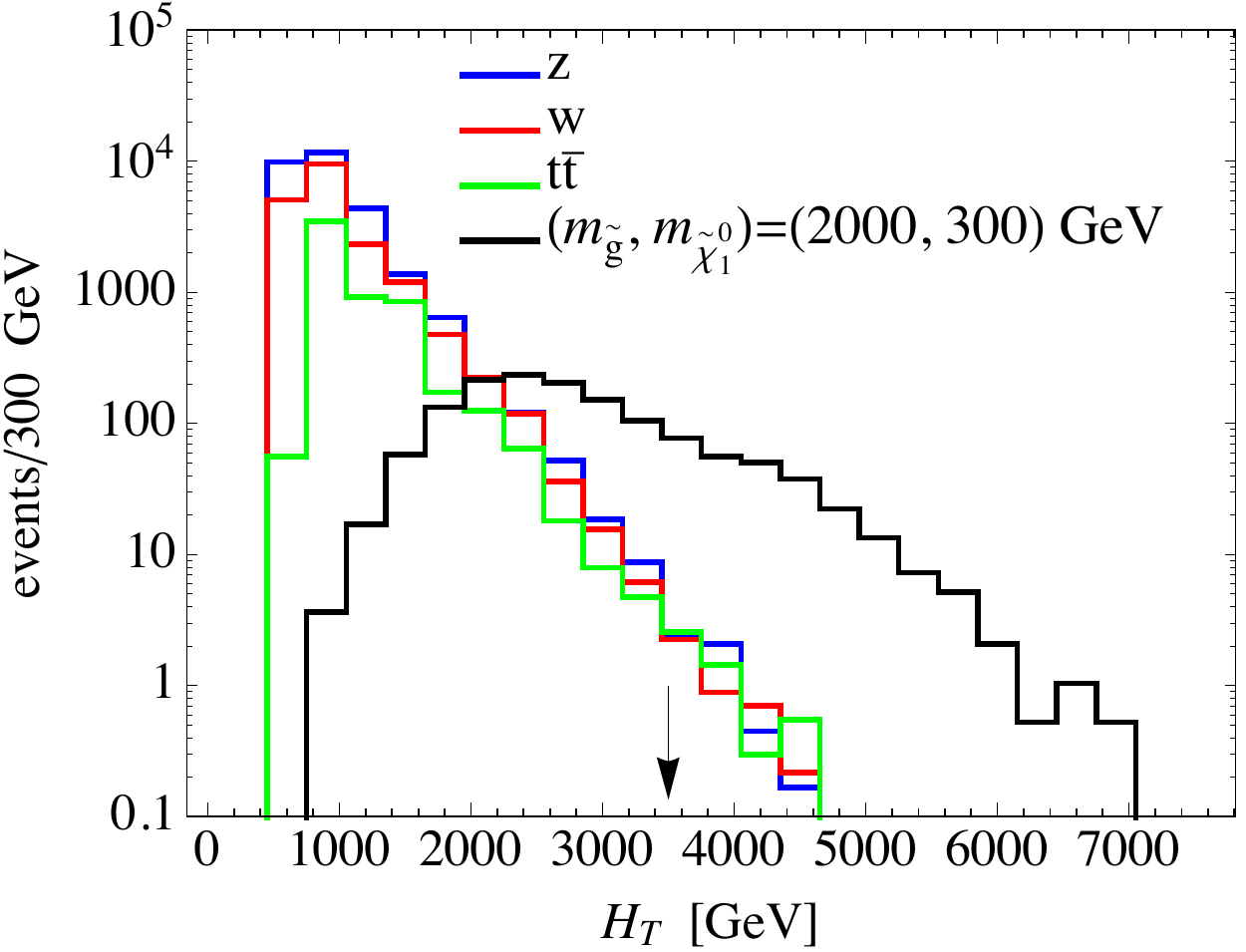}
\caption{\footnotesize The $H_T$ distribution after all cuts except for that on the $H_T$ are applied, at HL-LHC. In the left/right panel, the signal corresponds to the benchmark points in Eq.~(\ref{eq:BP1}). The arrows indicate the final cuts on $H_T$ for the chosen signal regions.
}
\label{fig:meffgluino}
\end{figure}

To present the qualitative features of our analyses, we choose two gluino benchmark points 
\begin{equation}
m(\tilde{g}, \tilde{\chi}^0) = (1020\ {\rm GeV}, 1000\ {\rm GeV})\ {\rm and}\  
(2000\ {\rm GeV}, 300\ {\rm GeV}). 
\label{eq:BP1}
\end{equation}
The first benchmark represents the nearly-degenerate (compressed) mass spectrum $-$ a challenging scenario for collider searches. The second set has a typical light LSP, implying large missing energy in the final state. 
In Table~\ref{tab:gluinocutflow}, we show the cut flow for two benchmark points, using our baseline selection cuts in addition to the optimal $\cancel{E}_T$ and $H_T$ cuts that we have found for each point. We also show sample kinematic distributions of $H_T$ for both signal and background in Figure~\ref{fig:meffgluino}, after all cuts except for the last cut are applied. 
We see that the value of $H_T$ is broadly peaked at $m_{\tilde{g}}$ as expected. The tail of the $H_T$ distribution extends further for a lighter LSP due to larger $\cancel{E}_T$.
With our preferred cuts, the significance is maximized.

\begin{figure}
\centering
\includegraphics[width=0.49 \textwidth ]{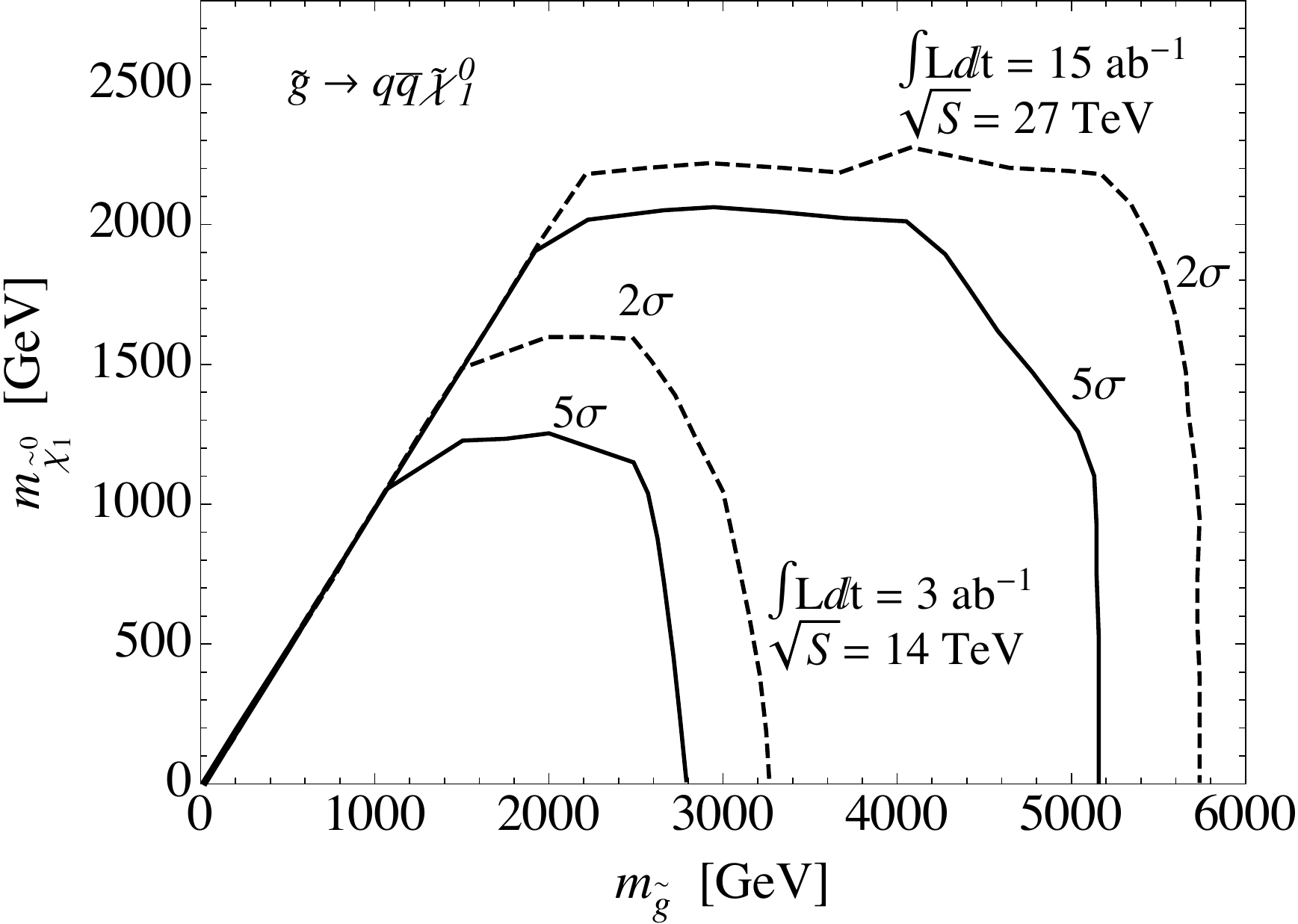}
\includegraphics[width=0.49 \textwidth ]{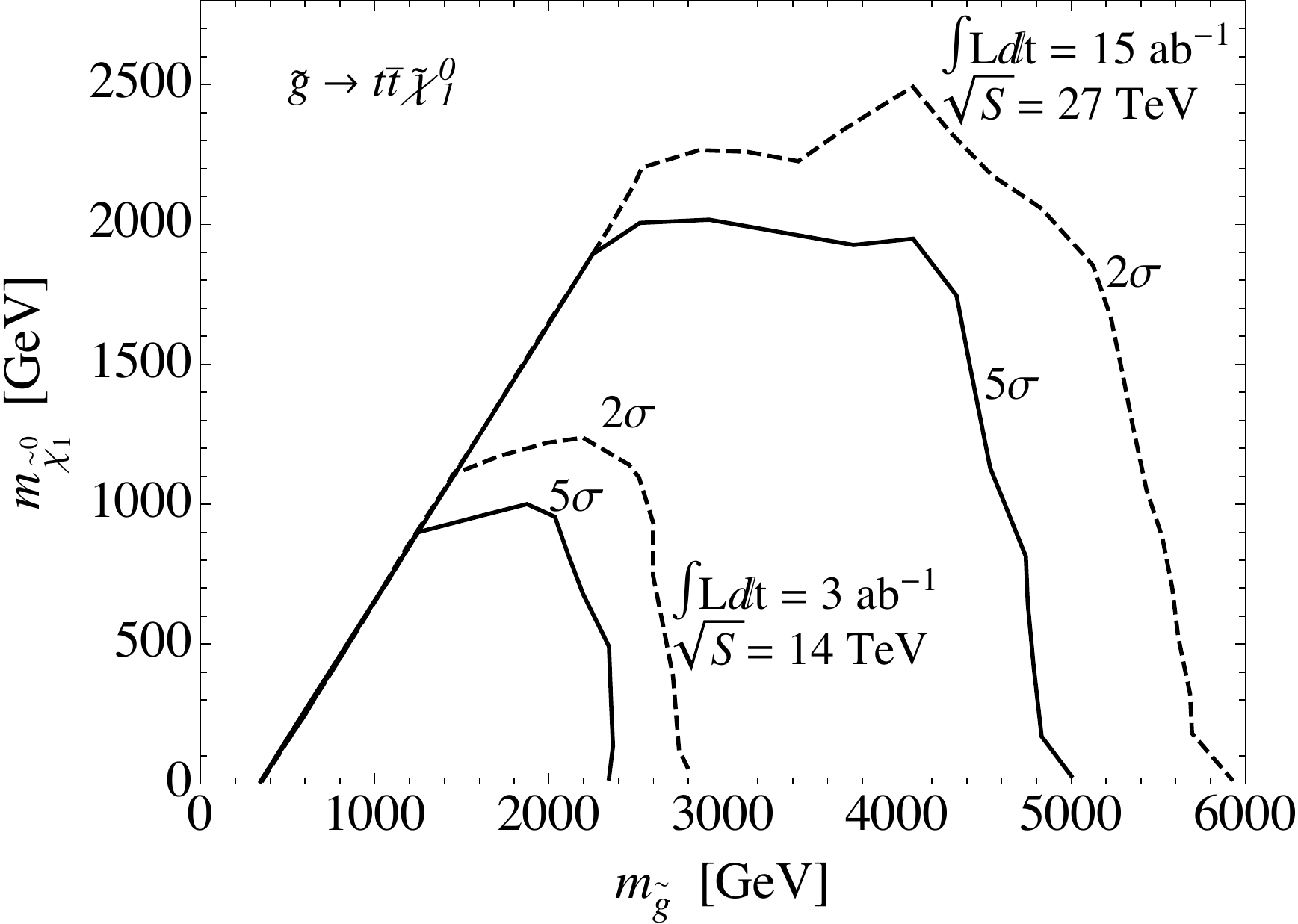}
\caption{\footnotesize 
The expected reach of LHC upgrades in probing gluinos decaying through off-shell squarks $\tilde{g} \to q \bar{q} \tilde{\chi}_1^0$ (left panel) and $\tilde{g} \to t \bar{t} \tilde{\chi}_1^0$ (right panel), in the gluino-LSP mass plane. The plots show the gluino mass reach at the LHC for 14 (27) TeV with 3 ab$^{-1}$ (15 ab$^{-1}$) of data. 
}
\label{fig:gluino}
\end{figure}

We show exclusion and discovery contours in Figure~\ref{fig:gluino} (left panel), indicating where the significance reaches $2\sigma$ (exclusion) and $5\sigma$ (discovery), respectively. We find that, for a massless LSP, a gluino of approximately 3.2 TeV can be probed by the HL-LHC with 3 ab$^{-1}$ of integrated luminosity. At 27 TeV with 15 ab$^{-1}$ of integrated luminosity, the exclusion (discovery) reach is roughly 5.7 (5.2) TeV.
We see that for a nearly-degenerate gluino and LSP, the exclusion (discovery) reach of HL-LHC and HE-LHC approximate 1.5 (1) TeV and 2.2 (2) TeV respectively. 
For comparison, a 100 TeV collider with 3 ab$^{-1}$ of data would be able to discover gluinos with the same decay topology up to 11 TeV, again assuming a massless LSP~\cite{Cohen:2013xda}.

\subsection{$\tilde{g} \to t \bar{t} \tilde{\chi}^0$}

For the simplified model with heavy flavor quarks in the final state, we still require four jets plus missing energy, as explained above, but now impose the additional demand that two of the jets be $b$-tagged, reducing the backgrounds of the previous subsection considerably. Instead, the main sources of background events are $t\bar{t}$, $t\bar{t}Z$ and $b\bar{b}Z$ production. We use the same event generation pipeline as for the $\tilde{g} \to q \bar{q} \tilde{\chi}^0$ topology.

We initially veto on leptons with $p_T$ above 10 GeV and $|\eta|$ below 2.5.
We then apply a set of baseline cuts 
\begin{equation}
N_\mathrm{jets}(p_T > 50~\mathrm{GeV},\, |\eta| <3.0)\geq 4, \qquad \cancel{E}_T > 100~\mathrm{GeV} 
\label{eq:cut2}
\end {equation}
For the HL-LHC (HE-LHC), we then search for events where the first two jets have $p_T > 200~(600)$ GeV and the third and fourth jets have $p_T > 80~(80)$ GeV.  Furthermore, we ask that at least two jets are b-tagged.
To discriminate signal events from QCD dijet events, we reject events with $S_T < 0.1$ where $S_T$ is the the transverse sphericity. We also reject events with $\Delta \phi(j, \cancel{E}_T) < \pi/18$ where $\Delta \phi(j, \cancel{E}_T)$is the transverse angle between $\cancel{E}_T$ and the closest jet, to reduce enhancement of $\cancel{E}_T$ due to mismeasurement of jet energies. In addition, we reduce background further by requiring $p_T(\mathrm{j_4})/H_T > 0.1$. After these baseline cuts, further cuts on $p_T(\mathrm{j_1})$ and $\cancel{E}_T / \sqrt{H_T}$ are optimized at each signal point to maximize the significance, which is calculated in the same fashion as above. 
For the HL-LHC (HE-LHC), we vary $p_T(\mathrm{j_1})$ in steps of 0.1 (0.1) TeV from 0.2 (0.4) up to 1 (1.6) TeV and $\cancel{E}_T / \sqrt{H_T}$ in steps of 1 (1) $\mathrm{GeV}^{1/2}$ from 10 (10) up to 27 (27) $\mathrm{GeV}^{1/2}$.

We show exclusion and discovery contours in Figure~\ref{fig:gluino} (right panel),
indicating where the significance reaches $2\sigma$ and $5\sigma$, respectively. For a massless LSP, a gluino of approximately 2.3 TeV can be discovered by the HL-LHC with 3 ab$^{-1}$ of integrated luminosity. At 27 TeV with 15 ab$^{-1}$ of integrated luminosity, we find that the exclusion (discovery) reach is roughly 5.7 (4.8) TeV. The latter result is somewhat weaker than that of~\cite{Baer:2018hpb}, with respect to which we find larger $b\bar{b}Z$ and $t\bar{t}$ backgrounds, though comparable to the reach in the direct gluino decay simplified model of the previous subsection. In both cases, the HE-LHC represents a significant gain over the HL-LHC in our ability to probe gluinos. A 100 TeV collider with 3 ab$^{-1}$ of integrated luminosity would be able to discover gluinos up to 6.4 TeV decaying as $\tilde{g} \to t \bar{t} \tilde{\chi}^0$, again assuming a massless LSP~\cite{Cohen:2013xda}.

\section{Stops}
\label{sec:stops}

The stop is arguably the most sought-after SUSY state because of its unique role in cancelling the quadratic sensitivity to the new physics scale in the Higgs boson mass corrections from top quarks \cite{Craig:2013cxa}. In addition, if the soft scalar masses are unified at some high scale, the stops are often the lightest accessible squarks. We thus study the reach of future colliders in probing stops. Because of its color and spin quantum numbers, the stop has a lower production cross section than the gluino. 
Including NLO + NLL QCD corrections \cite{Borschensky:2014cia}, for a stop mass of 1.5 TeV it is
$${\rm 0.4\ fb \ at\ 14\ TeV\ and\ 9.2\ fb\ at\ 27\ TeV}.$$
As in the previous section, the 27 TeV NLO + NLL K-factor has been estimated from the 33 TeV cross section and the rate is enhanced by a factor of 23 from 14 TeV to 27 TeV. We restrict ourselves to the case where $\tilde{t} \to t + \tilde{\chi}^0$ with 100\% branching fraction. Owing to the different possible decay modes of the tops, there are several possible final states, involving $b$-jets and missing energy with either 0, 1 or 2 leptons. Because of the $\approx 20\%$ leptonic branching fraction of the $W$ boson produced in each top decay, the final states with more leptons are more rare despite being cleaner to search for experimentally due to lower backgrounds. Guided by existing limits~\cite{Aaboud:2017ayj,Sirunyan:2017leh,Aaboud:2017aeu}, we choose to estimate the reach of an all-hadronic stop search with similar methodology to~\cite{ATL-PHYS-PUB-2013-011}. Our final state of interest is thus six jets and missing energy, including two $b$-tagged jets, and with two triplets of jets each forming tops. The background from SM processes in this state is dominated by $t\bar{t}$ and $t\bar{t}Z$ production, with smaller contributions from $t\bar{t}W$ and $V$ + jets which we neglect. We generate signal and background events using the same pipeline as for the gluino search above.

\begin{table*}
\centering
\begin{tabular}{|c|c|c|c|}
\hline
$m(\tilde{t}, \tilde{\chi}^0)$  & (500, 300) & (1500, 300)& Background\\
\hline \hline
Generator-level cuts & 662 & 0.40 & $2.92 \cdot 10^5$\\
Lepton veto & 473 & 0.31 & $1.44 \cdot 10^5$\\
Jet $p_T > (80, 80, 35, 35, 35, 35)$~GeV & 34.2 & 0.07 & $3.14 \cdot 10^3$\\
2 $b$-jets & 17.3& 0.05& $1.69 \cdot 10^3$ \\ 
$\Delta \phi(j_{1,2,3},\cancel{E}_T) > 0.2 \pi$ & 10.1 & 0.04 & 843 \\
Top reconstruction & 4.7 & 0.01 & 301\\
\hline \hline
$m_T(b) > 300~\mathrm{GeV}$ & 0.09 & -- & 0.61\\
$\cancel{E}_T > 300$~GeV & 0.03 & -- & 0.43 \\
\hline
$m_T(b) > 400~\mathrm{GeV}$ & -- & 0.005 & 0.15 \\
$\cancel{E}_T > 800$~GeV & -- & 0.002 & 0.002\\
\hline
\end{tabular}
\caption{Cut flow for the two stop benchmark points in Eq.~(\ref{eq:BP2}) at HL-LHC. At each step, the cross section remaining after the indicated cut is shown in fb. The baseline cuts are identical for both points.}
\label{tab:stopcutflow}
\end{table*}

To extract the signal, we begin by imposing the baseline cuts as in Eq.~(\ref{eq:cut}).
We once again reject events with leptons by requiring that signal events contain no electrons (muons) with $p_T$ above 20 (10) GeV and $|\eta|$ below 2.5 (2.4).
 We further require the leading two (next four) jets to have $p_T > 80\ (35)$ GeV and $|\eta| < 2.5$, with at least two $b$-tagged jets. Next, we require a minimum angular separation between the three leading jets and the $\cancel{E}_T$ of $\Delta \phi(j, \cancel{E}_T) > 0.2\pi$. We then reconstruct hadronic top quarks as follows~\cite{Aad:2012xqa}. The two closest jets in $\Delta R=\sqrt{\Delta \phi^2+\Delta \eta^2}$ are considered as a $W$ candidate from the top decay, and the next-closest jet to the $W$ candidate is combined to form a top candidate. Then, the same technique is applied to the remaining jets in the event, to extract the second top. If the mass of either top candidate is outside the region [80, 270] GeV, the event is rejected. Following the preceding selections, we optimize further cuts on $\cancel{E}_T$ over the interval [300, 1000]\,([300, 2500]) GeV, and the bottom transverse mass $m_T^b$ over the interval [200, 600]\,([200, 1500]) GeV, where the latter is constructed using the $b$-jet with the least angular separation from the $\cancel{E}_T$.

\begin{figure}
\centering
\includegraphics[width=0.49 \textwidth ]{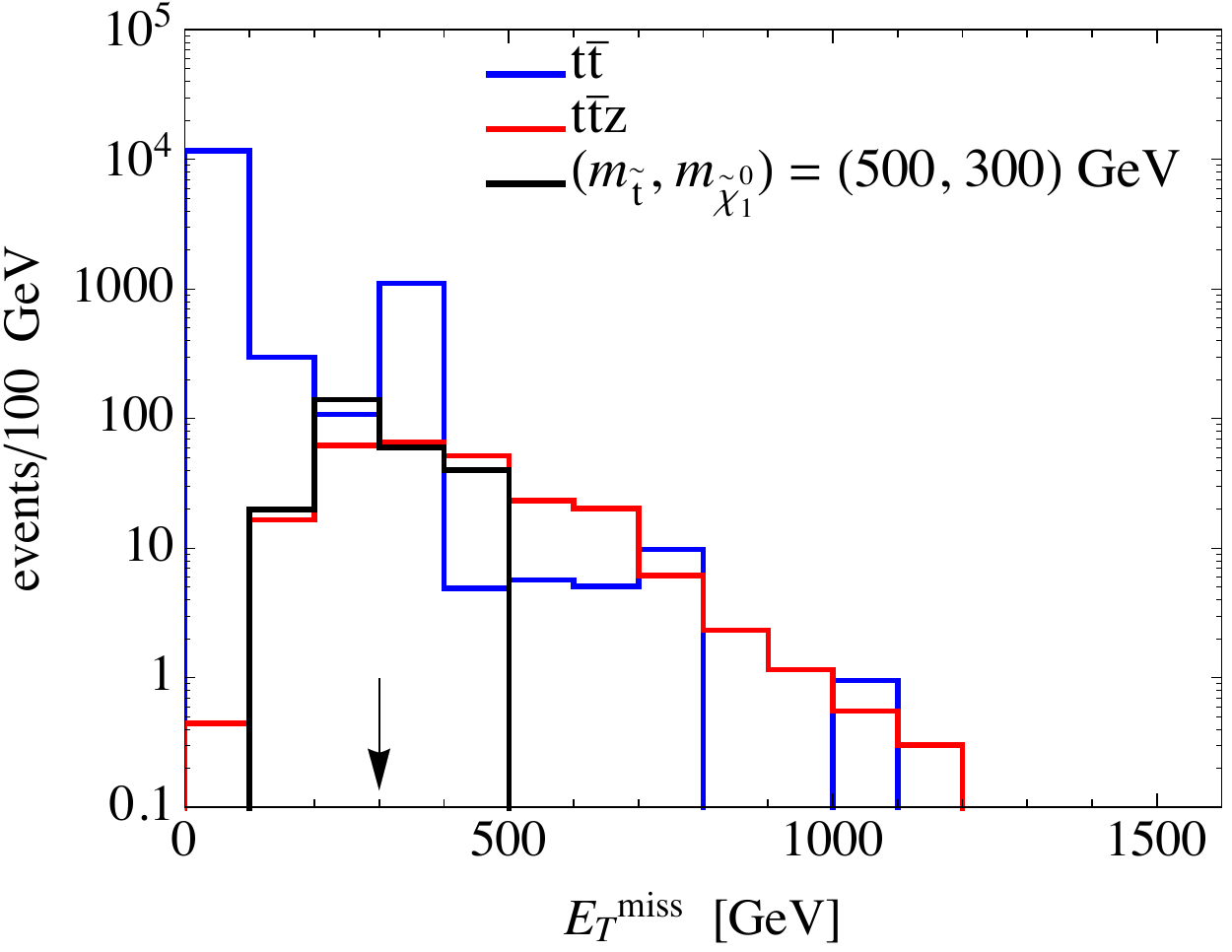}
\includegraphics[width=0.49 \textwidth ]{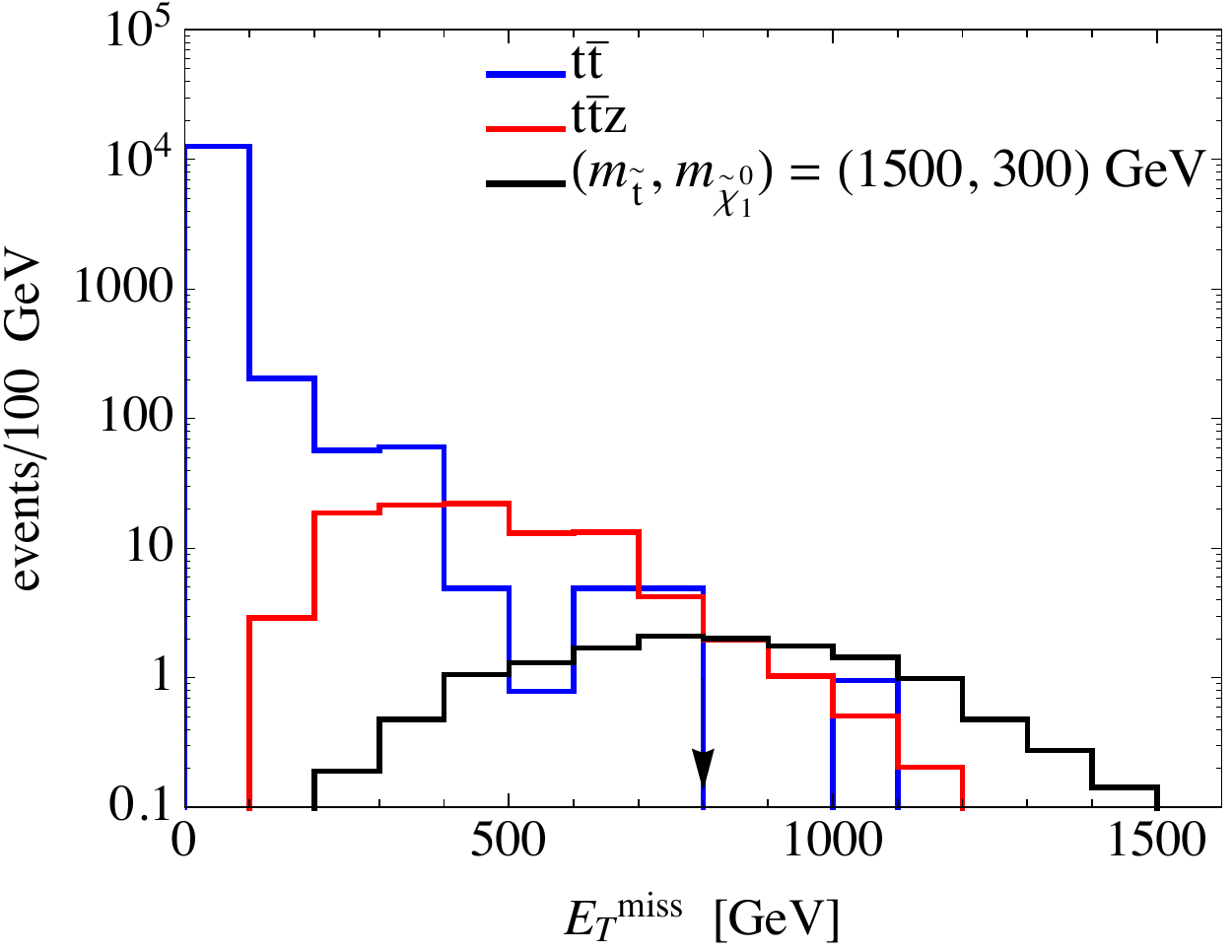}
\caption{\footnotesize The $\cancel{E}_T$ distribution after all cuts except for that on the $\cancel{E}_T$ are applied, at HL-LHC. In the left/right panel, the signal corresponds to the benchmark points in Eq.~(\ref{eq:BP2}). The arrows indicate the final cuts on $\cancel{E}_T$ for the chosen signal regions.}
\label{fig:metstop}
\end{figure}

To present the qualitative features of our analyses, we once again choose two benchmark points, with stop and LSP masses
\begin{equation}
m(\tilde{t}, \tilde{\chi}^0) = (500\ {\rm GeV}, 300\ {\rm GeV})\ {\rm and}\  
(1500\ {\rm GeV}, 300\ {\rm GeV}). 
\label{eq:BP2}
\end{equation}
Again, the first point represents the nearly-degenerate (compressed) mass spectrum--a challenging scenario for collider searches. The second set has a standard light LSP, yielding large missing energy. 
Table~\ref{tab:stopcutflow} shows the cut flows for two benchmark stop scenarios. Once again, the more compressed point benefits from looser cuts. Figure~\ref{fig:metstop} then shows the kinematic distributions after applying all cuts except for the optimized cut on $\cancel{E}_T$ for these benchmarks, demonstrating that at the expense of allowing more background, we achieve better significance for more compressed stops by retaining events with low $\cancel{E}_T$. It is primarily due to the optimization of our cuts that we are able to achieve greater sensitivity than the ATLAS study, which varied the $\cancel{E}_T$ and $m_T(b)$ cuts simultaneously with the stop mass. 

\begin{figure}
\centering
\includegraphics[width=0.6 \textwidth ]{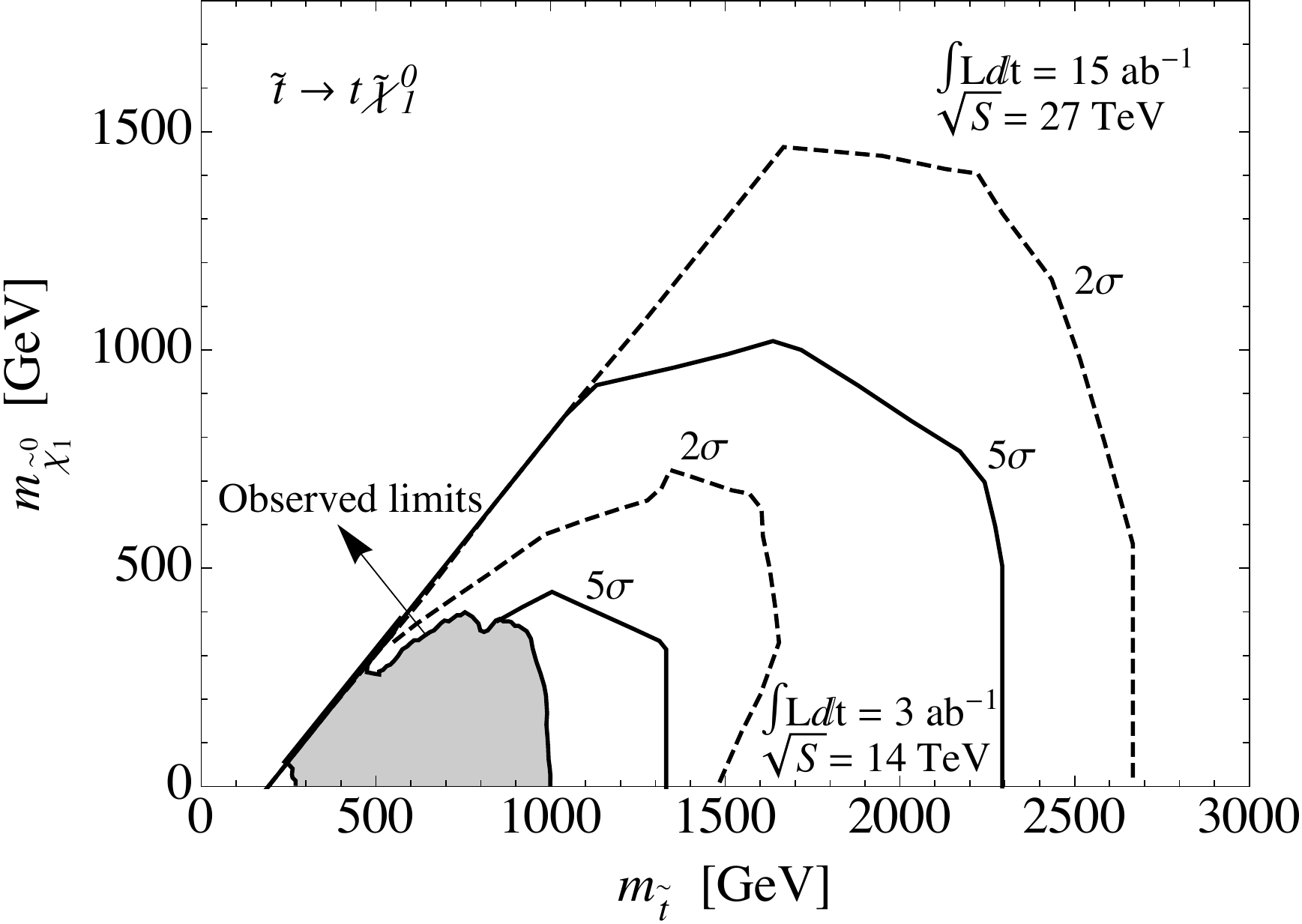}
\caption{\footnotesize 
The expected reach of LHC upgrades in probing stops decaying through $\tilde{t} \to t \tilde{\chi}_1^0$ in the stop-LSP mass plane. The plot shows the stop mass reach at 14 (27) TeV with 3 ab$^{-1}$ (15 ab$^{-1}$) of data. The shaded region indicates the current observed limits from the LHC.
}
\label{fig:stop}
\end{figure}

The results are shown in Figure~\ref{fig:stop}, demonstrating that for a massless LSP, a stop of approximately 1.5 TeV can be probed by the HL-LHC with 3 ab$^{-1}$ of integrated luminosity. At 27 TeV with 15 ab$^{-1}$ of integrated luminosity, the exclusion (discovery) reach is roughly 2.7 (2.3) TeV.
Unlike the gluino case, a direct comparison with an even higher potential collider is more difficult because of the challenges in reconstructing boosted top quarks. However, discovery of 5.5 TeV stops may be achieved at a 100 TeV collider with 3 ab$^{-1}$ of data using new top tagging techniques~\cite{Cohen:2014hxa}, or even 8 TeV stops with 30 ab$^{-1}$ of data~\cite{Mangano:2018mur}.

\section{Discussion and Conclusions}
\label{sec:concl}

In this work, we have evaluated the reach of the high-luminosity and high-energy upgrades of the LHC in searching for signals of strongly produced supersymmetry. Using simplified models based on production of gluinos and stops, two of the new particles most closely tied to the tuning of the electroweak scale in supersymmetry, we have adapted the existing experimental analyses, optimizing cuts to maximize the sensitivity to both conventional and compressed mass spectra. For gluinos, we find that the HL-LHC and HE-LHC have discovery reach up to masses of 2.8 and 5.2 TeV, respectively. Similarly, our study shows that the HL-LHC and HE-LHC can discover stops with masses up to 1.3 and 2.3 TeV.

Our results for HL-LHC are slightly better than those in existing experimental projections. The main reason for the improvement is that we have adjusted two cuts in each of our analyses to aid in discerning signal from background, which is expected to yield better reach than applying a fixed set of cuts for all potential super-partner masses, or even than performing a one-dimensional optimization. Conversely, it is possible that changing more cuts or even more sophisticated search strategies could improve the results that we have found. As an example, a recent ATLAS search for the same gluino simplified model as we have considered excluded gluinos at 2.0 TeV with 36 fb$^{-1}$ of integrated luminosity. This analysis used a combination of an $m_{\mathrm{eff}}$-based search as we have performed but with additional cuts, as well as signal regions constructed using the recursive jigsaw reconstruction technique. Extrapolating this gluino exclusion up to the HE-LHC by scaling the gluon parton distribution function would yield an expected exclusion of over 7 TeV, suggesting that further refinements could be made to our search. Nevertheless, our results provide a useful estimate of the practical reach that an upgraded LHC could achieve in searching for stops and gluinos. Our procedure should serve as useful guidance for future analyses.

Finally, we comment that while we have considered only simplified models with two particles, in a more complicated scenario it would be more involved to determine the reach of a future collider. For instance, intermediate electroweak gauginos between the strongly produced particle and the LSP would give final states with leptons, which we have not considered, and a variety of different cascade decays would be possible depending on the super-partner mass splittings and mixings.

As the LHC continues to acquire more data, it is important to assess the sensitivity of its potential successors to BSM theories. Here, we have studied the reach of the high luminosity and high energy upgrades of the LHC in probing gluinos and stops in supersymmetry. Further work is warranted to continue the exploration of the ability of these machines and beyond to explore new physics at the energy frontier.

\section*{Acknowledgements}
We thank M.L.~Mangano for encouragement to carry out these analyses, and H.~Baer for discussions. We also thank the Aspen Center for Physics, which is supported by National Science Foundation grant PHY-1607611, for its hospitality during the completion of this work. We also thank the Galileo Galilei Institute for Theoretical Physics for the hospitality and the INFN for partial support during the completion of this work.
This work was supported in part by the Department of Energy under Grants No.~DE-FG02- 95ER40896 and DE-SC0015634, and in part by PITT PACC.

\section*{References}

\bibliography{susy}

\end{document}